\documentstyle[preprint,aps]{revtex}
\begin{document}
\draft
\title{Long-time  dynamics of the infinite-temperature Heisenberg
magnet
\thanks{To be published in PRB, Feb. 1994}}
\author
{Michael Chertkov $^*$ and Igor Kolokolov $^\dagger $}
\address{ $^*$\ Physics Department, Weizmann Institute of Science,
Rehovot 76100, Israel\\ $^\dagger $\ Budker Institute for Nuclear Physics,
Novosibirsk 630 090, Russia
\\ and INFN, Sezione di Milano, 20133 Milano, via Celoria 16, Italy}
\date{\today}
\maketitle
\begin{abstract}
Infinite-temperature long-time dynamics of Heisenberg model
${\bf\hat{H}}=-\frac{1}{2}\sum_{i,j}J_{ij}
\hat{\vec{S}}_{i}\hat{\vec{S}}_{j}$ is investigated. It is shown that
the quantum spin pair-correlator is equal to the correlator of classically
evaluated vector field
averaged over the initial conditions
with respect to the gaussian measure.
In the continious limit case the scaling estimations
allow one to find one-point correlator  that turns out to be
$C(\vec{r}=0;t)\propto const \times t^{-6/7}$.
All results are obtained by straightforward procedure without any
assumptions of the phenomenological character.
\end{abstract}
\pacs{PACS number 75.10.J, 02.30.C}
\overfullrule=0pt
\section*{Introduction}
Because of the boundedness of the spin operators the problem of
 infinite-temperature
 dynamics of quantum Heisenberg model is well defined. For the equilibrium case
 it
reduces to the computation of correlators like follows
\begin{equation}
C^{\alpha\beta} (\vec{r};t)=Tr[\exp^{i t{\bf\hat{H}}} \hat{S}^{\alpha}_{i}
\exp^{-it{\bf\hat{H}}}
\hat{S}^{\beta}_{j}].
\label{1}
\end{equation}
Here ${\bf\hat{H}}$ is the Heisenberg exchange Hamiltonian
\begin{equation}
{\bf\hat{H}}=-\frac{1}{2}\sum_{i,j}J_{ij}\hat{\vec{S}}_{i}\hat{\vec{S}}_{j} ,
\label{2}
\end{equation}
$\hat{S}^{\alpha}_{i}$ is the $\alpha$-th spin operator component on the
lattice
 site
$i$, and $\vec{r}=\vec{r}_{i}-\vec{r}_{j}$ is the distance between lattice
sites $i$ and $j$.
The thermodynamic limit case ($\sum_{i} 1=\infty$) is supposed.

 For fixed $\vec{r}$ and
$t\rightarrow 0$ the correlator $C^{\alpha \beta}=\delta^{\alpha\beta}
C(\vec{r};t)$ can be
 calculated
by the direct temporal expansion of the evolution operator
$\exp(it{\bf\hat{H}})$ (see the initial papers \cite{Col,McFad,Benn},  the
review
 of the results obtained by this approach
 in the work \cite{Labr} and the more recent paper
 \cite{Zob}).
Attempts to investigate the long-time dynamics starting from the expression
 (\ref{1})
have been done by a number of authors \cite{Benn,Res,Blum,TerH}.
The main
 tool
used in these papers was the correlators decoupling method.
 However, it can  only work in
the case when the dynamics consists mainly of well-defined propagating
 excitations, but this is not correct in our case and may lead
wrongly to obtain whatever is supposed. So, the most popular theory of
such kind
so-called spin-diffusion theory \cite{Benn,Blum} gives the spin
auto-correlation
function in the form $C(t)\propto t^{-3/2}$.

In the paper \cite{Kol1} the functional integral
 representation for (\ref{1})
 has been
derived. The small time $t$ but large
$\vec{r}^{2}t$ limit has been calculated
 with the
use of this representation.
In the present paper
we show that {\em under some controled
 assumptions} it is possible to extend this approach to
long-time limit case ($Jt\gg 1$) .
The problem reduces to the averaging of some classical equation over initial
conditions
 (see eqs(\ref{27},\ref{28},\ref{29})).
 In the continious limit case scaling estimates
allow one to find one-point correlator  (see the last section) which turns out
to be
\begin{equation}
C(t)\propto const \times t^{-\frac{6}{7}} ,
\label{3}
\end{equation}
and to prove the validity of our approximation at large enough time.

\section*{ Functional representation}
We start from the Hubbard-Stratonovich transformations of the generating
 functional of
the spin correlators
\begin{eqnarray}
&  &\bf{Q}[\vec{h}]  = Tr[\bf T_{c}\exp(i\int_{c}\it{d}
t({\bf\hat{H}}+\vec{h}_{i}\hat{\vec{S}}_{i}))] = \nonumber \\
 & = & N\int \it{D}\vec{\varphi}^{(1)} \it{D}\vec{\varphi}^{(2)}
\exp(\frac{i}{2}\sum_{i,j}\int_{c}\it{d}\acute{t}
 \vec{\varphi}_{i}(\acute{t})J_{ij}^{-1}\vec{\varphi}_{j}(\acute{t}))
\times \nonumber \\
& \times & \prod_{i}\bf T_{c} \exp(i\int_{c}\it{d} \acute{t} (
 \vec{\varphi}_{i}(\acute{t})+\vec{h}_{i}(\acute{t}))\hat{\vec{S}}_{i})\},
\label{4}
\end{eqnarray}
where $N$ is a normalization factor.
The symbol $\bf T_{c}$ indicates the ordering of operators along the contour
$c$ in
 the $t$-plane and
consists from two oppositely directed branches that pass
in the vicinity of real axes \cite{Kol1,Kol2,Kol3,Kol4}.
$\vec{h}_{i}(t)$ is the auxiliary external field.
The indices $1,2$ of the vector-valued fields indicate
the  upper and the lower branch of the $c$-contour respectively.
If we define the external field as
\begin{eqnarray}
\vec{h}_{j}^{(1)} & = & \vec{a}_{j} \delta (t) +\vec{b}_{j} \delta (t-T),
 \nonumber \\
\vec{h}_{j}^{(2)} & = & -\vec{a}_{j} \delta (t) -\vec{b}_{j} \delta (t-T),
\label{5}
\end{eqnarray}
then the correlator (\ref{1}) is
\begin{equation}
C^{\alpha \beta}(\vec{r};t)= \left.
-\frac{1}{4} \frac{\partial^{2} \bf{Q}(\vec{h})}{\partial a^{\alpha}_{i}
\partial b^{\beta}_{j}} \right|_{\vec{a}=\vec{b}=0} .
\label{6}
\end{equation}
Making the translation $\vec{\varphi} \rightarrow \vec{\varphi}-\vec{h}$ and
 neglecting  the terms
which do not contribute to (\ref{6}) we get the expression
\begin{eqnarray}
&  & \bf{Q}[\vec{h}] =
 \int \it{D}\vec{\varphi}^{(1)} \it{D}\vec{\varphi}^{(2)} \prod_{i}
Tr(\hat{A}_{i}(-\infty,+\infty)) \times   \nonumber \\
&\times & \exp\{\frac{i}{2}\sum_{i,j}\int_{c}\it{d}\acute{t}
(\vec{\varphi}^{(1)}_{i}
(\acute{t})J_{ij}^{-1}\vec{\varphi}^{(1)}_{j}(\acute{t})-\nonumber \\
&-&\vec{\varphi}^{(2)}_{i}(\acute{t})J_{ij}^{-1}\vec{\varphi}^{(2)}_{j}(\acute{t}))+
\vec{b}_{i}J_{ij}^{
 -1} (
\vec{\varphi}^{(1)}_{j}(t)+\vec{\varphi}^{(2)}_{j}(t))+\nonumber \\
&+ & \vec{a}_{i}J_{ij}^{-1} (
\vec{\varphi}^{(1)}_{j}(0)+\vec{\varphi}^{(2)}_{j}(0))\}.
\label{7}
\end{eqnarray}
Here the operator $\hat{A}(-\infty,t)$ for a given lattice site is defined as
\begin{eqnarray}
& \hat{A} & (-\infty,t)=
 {\bf T}\exp(i \int_{-\infty}^{t}
\vec{\varphi}^{(1)}_{i}(\acute{t})\hat{\vec{S}} \it{d} \acute{t})
\times \nonumber \\
& \times &{\bf \tilde{T}} \exp(-i \int_{-\infty}^{t}
\vec{\varphi}^{(2)}_{i}(\acute{t})\hat{\vec{S}} \it{d}
 \acute{t}),
\label{8}
\end{eqnarray}
and ${\bf \tilde{T}}$ denotes anti-chronological ordering.
This operator is determined by the
differential equation
\begin{eqnarray}
& - & i \frac{\it{d} \hat{A}(-\infty,t)}{\it{d}t}=\nonumber \\
& = &(\vec{\varphi}^{(1)}_{i}(t)\hat{\vec{S}})\hat{A}(-\infty,t)-
\hat{A}(-\infty,t)(\vec{\varphi}^{(2)}_{i}(t)\hat{\vec{S}}) ,\nonumber \\
\label{9}
\end{eqnarray}
and by the initial condition $\hat{A}(-\infty,-\infty)=1$.
Let us perform the ansatz ( see also \cite{Kol1,Kol2} )
\begin{equation}
\hat{A}(t)=\exp(i
\int_{-\infty}^{t}(\vec{\rho}^{(1)}(\acute{t})-\vec{\rho}^{(2)}(\acute{t}))\hat{\vec{S}} \it{d}\acute{t}),
\label{10}
\end{equation}
where $\vec{\rho}^{(1)}(t),\vec{\rho}^{(2)}(t)$ are some new vector fields.
Differentiating this
 equation with respect
to time  we obtain
\begin{eqnarray}
& - & i\frac{\it{d} \hat{A}(-\infty,t)}{\it{d}t}=\nonumber \\
& = & \int_{0}^{1}\it{d}\tau \exp(i\tau
 \vec{\zeta}\hat{\vec{S}})(\vec{\rho}^{(1)}-\vec{\rho}^{(2)})
\hat{\vec{S}}\exp(i(1-\tau ) \vec{\zeta}\hat{\vec{S}}) =
\nonumber \\
& = & \hat{U}_{1}(t)\hat{A}(t)-\hat{A}(t)\hat{U}_{2}(t).
\label{11}
\end{eqnarray}
Here we introduce the notations of the following type
\begin{eqnarray}
\vec{\zeta}(t)=\int_{-\infty}^{t}\it{d}\acute{t} (\vec{\rho}^{(1)}
(\acute{t})-\vec{\rho}^{(2)} (\acute{t})),
\label{12} \\
\hat{U}_{1}(t)=\int_{0}^{1}\it{d}\tau \exp(i\tau
\vec{\zeta}\hat{\vec{S}})\vec{\rho}^{(1)}
 \hat{\vec{S}}\exp(-i\tau \vec{\zeta}\hat{\vec{S}}),
\nonumber \\
\hat{U}_{2}(t)=\int_{0}^{1}\it{d}\tau \exp(-i\tau
\vec{\zeta}\hat{\vec{S}})\vec{\rho}^{(2)}
 \hat{\vec{S}}\exp(i\tau ) \vec{\zeta}\hat{\vec{S}}).
\label{13}
\end{eqnarray}
Noting that the operator of the spin rotations  appeares in the integrands of
(\ref{13})  ,
we get immediately
\begin{eqnarray}
\hat{U}_{1}(t)=\vec{\eta}_{1}\hat{\vec{S}} \ , \  \  \
\hat{U}_{2}(t)=\vec{\eta}_{2}\hat{\vec{S}}
 ,\nonumber \\
\vec{\eta}_{1}=\vec{\rho}^{(1)}

\frac{\sin{|\vec{\zeta}|}}{|\vec{\zeta}|}+\vec{\zeta}\frac{\vec{(\zeta}\vec{\rho}^{(1)} )}
{\vec{\zeta}^{2}}(1-\frac{\sin{|\vec{\zeta}|}}{|\vec{\zeta}|})+
\nonumber \\
+\frac{1}{\vec{\zeta}^{2}}
[\vec{\zeta}\times \vec{\rho}^{(1)} ](1-\cos{|\vec{\zeta}|}),\nonumber \\
\vec{\eta}_{2}=\vec{\rho}^{(2)}

\frac{\sin{|\vec{\zeta}|}}{|\vec{\zeta}|}+\vec{\zeta}\frac{\vec{(\zeta}\vec{\rho}^{(2)} )}
{\vec{\zeta}^{2}}(1-\frac{\sin{|\vec{\zeta}|}}{|\vec{\zeta}|})-
\nonumber \\
-\frac{1}{\vec{\zeta}^{2}}
[\vec{\zeta}\times \vec{\rho}^{(2)} ](1-\cos{|\vec{\zeta}|}),
\label{14}
\end{eqnarray}
and
\begin{eqnarray}
& &
Tr[\hat{A}(-\infty,+\infty)]=\exp(g_{s}(|\int_{-\infty}^{+\infty}(\vec{\rho}^{(1)}-\vec{\rho}^{(2)})\it{
 d}t|)),
\nonumber \\
& & g_{s}(x)=\ln(\frac{\sin((S+1/2)x)}{\sin(x/2)}),
\label{15}
\end{eqnarray}
where $S$ is the spin magnitude.
Comparing eq. (\ref{9}) with eqs. (\ref{11},\ref{12},\ref{13},\ref{14}) we
conclude that the parameterization
\begin{equation}
\vec{\varphi}^{(1)}(t)=\vec{\eta}_{1}(t) \ , \  \
\vec{\varphi}^{(2)}(t)=\vec{\eta}_{2}(t)
\label{16}
\end{equation}
 gives the explicite form of eq.(\ref{15}) for
 $Tr[\hat{A}(-\infty,+\infty)]$.
It is more convenient to rewrite the equalities (\ref{14}) in terms of the
$\vec{\psi}$, $\vec{\zeta}$ fields
\begin{eqnarray}
& & \vec{\varphi}^{(1)}+\vec{\varphi}^{(2)}=\vec{\psi}, \nonumber \\
& & \vec{\varphi}^{(1)}-\vec{\varphi}^{(2)}= \dot{\vec{\zeta}}
\frac{\sin{|\vec{\zeta}|}}{|\vec{\zeta}|}+
\vec{\zeta}\frac{\vec{(\zeta}\dot{\vec{\zeta}} )}
{\vec{\zeta}^{2}}(1-\frac{\sin{|\vec{\zeta}|}}{|\vec{\zeta}|})-
\nonumber \\
& & -\frac{(1-\cos{|\vec{\zeta}|})}{\vec{|\zeta}|\sin{|\vec{\zeta}|} }
\{[\vec{\zeta}\times \dot{\vec{\zeta}} ]
+\frac{1}{\vec{\zeta}^{2}}(1-\cos{|\vec{\zeta}|})[\vec{\zeta}
\times [\vec{\zeta} \times \dot{\vec{\zeta}}]]\},\nonumber\\
\label{17} \\
& & \vec{\zeta}(-\infty)=0,
\label{18}
\end{eqnarray}
where
\begin{equation}
\vec{\psi}=\vec{\rho}^{(1)}+\vec{\rho}^{(2)} \ , \
\vec{\zeta}=\int_{-\infty}^{t}(\vec{\rho}^{(1)}(\acute{t})-\vec{\rho}^{(2)}(\acute{t}))\it{d}\acute{t} .
\label{19}
\end{equation}
We can consider $ \vec{\psi}, \vec{\zeta}$  as the new
integration variables in eq. (\ref{7}) . This allow us to write down
explicit functional representation for the generating
functional $\bf{Q}[\vec{h}]$.
Let us note, that the essential difference with \cite{Kol1} is that the
transformations (\ref{17})
  are nonlinear only in the
subspace of fields configurations .  Indeed, one of the new integration
variables, the
 field $\vec{\psi}$ , is
connected with the old integrations variables $\vec{\varphi}^{(1)}$ and
$\vec{\varphi}^{(2)}$  linearly.
The Jacobian of the transformation
 (\ref{18}) has the "ultralocal"
form
\begin{equation}
J=const \times \prod_{t}\{1+\frac{6}{|\vec{\zeta}|}
\tan{|\vec{\zeta}|/2}(\frac{2 \tan{|\vec{\zeta}|/2}}{
|\vec{\zeta}|}-1)\}
\label{20}
\end{equation}

\section*{Effective equation of motion}
The effective action that is obtained after the substitution
(\ref{15},\ref{18}) into
(\ref{7}) is rather complicated and some approximations are necessary.
It seems to be reasonable to assume that for the long-time dynamics the
relevant
fields configurations obey the inequality
\begin{equation}
|\vec{\zeta}(t)|=|\int_{-\infty}^{t}(\vec{\rho}^{(1)} (\acute{t})-
\vec{\rho}^{(2)}(\acute{t}))\it{d}\acute{t}|\ll 1 .
\label{21}
\end{equation}
 This assumption is based on some facts. First of all, the action does
not contain even with respect to $\dot{\vec{\zeta}}(t)$ terms. It leads ,
in particular, to the zero value of all the correlators containing field
 $\dot{\vec{\zeta}}(t)$ at least once. Secondly, all physical ( not
containing the field  $\dot{\vec{\zeta}}(t)$ ) correlators must be close to
zero
as t goes to $\infty$ due to absence of the spreading excitations at infinite
temperature. But in any case, to check the validity of assumption
(\ref{21}) we have to estimate the neglected
terms (see the last paragraph of the next section).

Expanding the right-hand side of (\ref{17}) in series of
 $\vec{\zeta}(t)$ and keeping only the first nontrivial term,
we write down the approximate form of the map (\ref{17}-\ref{20})
\begin{eqnarray}
\vec{\varphi}^{(1)}+\vec{\varphi}^{(2)}=\vec{\psi}, \nonumber \\
\vec{\varphi}^{(1)}-\vec{\varphi}^{(2)}\approx \dot{\vec{\zeta}}+\frac{1}{2}
[\vec{\psi} \times\vec{\zeta}]
\label{22} \\
J\approx 1
\label{23}
\end{eqnarray}
The approximation for $Tr[\hat{A}(-\infty,+\infty)]$ is
\begin{equation}
Tr[\hat{A}(-\infty,+\infty)]\approx \exp(-\frac{D}{2}\vec{\zeta}^2(+\infty)),
\label{24}
\end{equation}
where $D=|\ddot{g}_{s}(0)|$ ( see (\ref{15}) ). Substituting (\ref{22}) into
(\ref{7}) and introducing the field $\vec{\phi}(t)$
\begin{equation}
\vec{\phi}_{i}=2J_{ij} \vec{\psi}_{j},
\label{25}
\end{equation}
that directly corresponds to the spin on the lattice site $i$,
we note that the $\it{D}\vec{\zeta}$ integration can be easily performed.
The classical equation of motion for the field $\vec{\phi}$
averaged over the "initial" data at the far future is obtained
\begin{eqnarray}
Q[\vec{a},\vec{b}]=\int \it{D}\vec{\phi}\exp\{-\frac{1}{2D}\sum_{i}
\vec{\phi}_{i}^{2}(+\infty)-\nonumber \\
-2\vec{a}_{i}\vec{\phi}_{i}(t)-
2\vec{b}_{i}\vec{\phi}_{i}(0)\}\prod_{i;t}\delta(\dot{\vec{\phi}}_{i}-
\sum_{j}J_{ij} [\vec{\phi}_{i}\times \vec{\phi}_{j}])
\label{26}
\end{eqnarray}
(Fourier-like integration with respect to the field $\vec{\zeta}(\acute{t})$
  in an intermediate time moment gives the
$\delta$-functions product in eq.(\ref{26}), while
the gaussian integration with respect to $\vec{\zeta}(+\infty)$ produces
probability distribution for $\vec{\phi}(+\infty)$).
It is important to note that the $\vec{\phi}_{i}^{2}$ term defining the weight
of
 averaging is the integral of motion of classical equations given in
(\ref{26}).  Let us mention that
from the ergodicity consideration it is natural to obtain the averaging over
the distribution
of the motion integral  that is the only non-stochastical variable in the
system.
Thus in our approximation the quantum spin correlator (\ref{1})
is equal to the correlator of the classically evaluated vector field
$\vec{\phi}_{i}(t)$
\begin{equation}
\dot{\vec{\phi}}_{i}=
\sum_{j}J_{ij} [\vec{\phi}_{i}\times \vec{\phi}_{j}]
\label{27}
\end{equation}
averaged over the initial conditions
\begin{equation}
\vec{\phi}_{i}(0)=\vec{p}_{i}
\label{28}
\end{equation}
with respect to the gaussian measure
\begin{equation}
\prod_{i}\it{d}\vec{p}_{i}\exp\{-\frac{1}{2D}\sum_{i}
\vec{p}_{i}^{2}\}.
\label{29}
\end{equation}
Here the conservation of the phase space volume element
$\it{d}\vec{\phi}_{i}$ during the evolution (\ref{27}) has been taken into
account. Note that eq.(\ref{27}) is only the spin-operator equation of motion
with the changing
$\hat{\vec{S}}$ to the classical evaluated field $\vec{\phi}$.

\section*{ Scaling estimations}
It is natural to suppose that the long-time evolution is determined by the
long-wavelength fluctuations. Thus we can use the continium version of
(\ref{27})
\begin{equation}
\dot{\vec{\phi}}(\vec{r};t)=
\alpha [\Delta\vec{\phi} \times \vec{\phi}]
\label{30}
\end{equation}
and (\ref{29})
\begin{equation}
\it{D}\vec{p}(\vec{r})\exp\{-\frac{1}{2\tilde{D}}\int \it{d}^{3}r
\vec{p}^{2}(\vec{r})\}.
\label{31}
\end{equation}
Here $\tilde{D}=a^{3}D$, $a$ is the lattice spacing, and $\alpha$ is defined in
terms of the Fourier transform $J(\vec{k})$ of the exchange integral
$J_{ij}=J(\vec{r}_{i}-\vec{r}_{j})$ as follows
\begin{equation}
J(\vec{k})\approx J(0)-w_{ex}(ak)^{2} \ , \ \ \alpha=w_{ex} a^{7/2}.
\label{32}
\end{equation}

The eqs. (\ref{31},\ref{32}) can be studied in principle with the use of
Wyld diagram technique \cite{Lv}. However, all the terms of the
perturbation theory  suffer from infrared singularities. On the other
hand, in the infrared region the scaling arguments can work. Indeed, the
equation (\ref{30}) is invariant with respect to the following continious
set of the scale-transformations group
\begin{eqnarray}
\vec{r} \rightarrow \lambda \vec{r} \  \ , \ \ t \rightarrow \lambda^{\beta} t,
\nonumber \\
\vec{\phi}(\vec{r};t) \rightarrow \lambda^{\beta -2} \vec{\phi}(\lambda\vec{r};
\lambda^{\beta} t).
\label{33}
\end{eqnarray}
Here $\beta$ is an arbitrary real number. The quantity
\begin{equation}
K(\vec{r};t)=\vec{\phi}(\vec{r};t)\vec{p}(\vec{r})
\label{34}
\end{equation}
obtained after averaging the desired one-lattice site spin-spin correlator
transforms as
\begin{equation}
K(\vec{r};t)\rightarrow \lambda^{2(\beta-2)}K(\lambda\vec{r};\lambda^{\beta} t)
\label{35}
\end{equation}
The requirement for the weight of averaging to be scaling invariant gives us
the unique value of $\beta$
\begin{equation}
\beta = \frac{7}{2}
\label{36}
\end{equation}
The invariance (\ref{33}) means that if some initial conditions
$\vec{\phi}(\vec{r};0)=\vec{p}(\vec{r})$ transforms to
$\lambda^{\beta-2}\vec{p}(\lambda\vec{r})$
than for any further moment $\vec{\phi}(\vec{r};t)$
transforms by (\ref{33}). For $\beta =7/2$ all points on each orbit generated
in the functional phase space by the scaling group (\ref{33}) have equal
probabilities. Thus the correlator $C(\vec{r};t)=<K(\vec{r};t)>$ should be
invariant
with respect to the transformations (\ref{33},\ref{36})
\begin{equation}
C(\vec{r};t)=\lambda^{3}C(\lambda \vec{r};\lambda^{7/2}t).
\label{37}
\end{equation}
Consequently, it has the form
\begin{equation}
C(\vec{r};t)=t^{-6/7}f(r/t^{2/7})
\label{38}
\end{equation}
and the one-point correlator finally is
\begin{equation}
C(t)\equiv C(\vec{r}=0;t)=const \times t^{-6/7}
\label{39}
\end{equation}

To estimate the contribution of the neglected terms let us
return to the functional integral with respect to
$\vec{\psi},\vec{\zeta}$ fields from the substitution (\ref{22}).
The suitable action is invariant with respect to
 the following
scaling transformation
\begin{eqnarray}
\vec{r} \rightarrow \lambda \vec{r} \   ,  \ t \rightarrow \lambda^{7/2} t \ ,
\  \
\vec{\psi}(\vec{r};t) \rightarrow \lambda^{3/2} \vec{\psi}(\lambda\vec{r};
\lambda^{7/2} t) ,
\nonumber \\
\vec{\zeta}(\vec{r};t) \rightarrow \lambda^{3/2} \vec{\zeta}(\lambda\vec{r};
\lambda^{7/2} t).
\label{40}
\end{eqnarray}
So, we can compute scaling indices for all the corrections to the correlator
$C(\vec{r};t)$. For the first nontrivial correction arising from the
non-linearities of the third
order non-linearities in (\ref{18}) with respect to the small parameter
(\ref{21}) we obtain
\begin{equation}
C^{(3)}(\vec{r};t)= t^{-12/7} f^{(1)}(r/t^{2/7})
\label{41}
\end{equation}
and it can be neglected  comparing with eq.(\ref{38}) in the limit
$t\rightarrow \infty$ and especially for $r=0$.
Every next order correction to the correlator gives an additional
 factor $t^{-3/7}$ and does not affect the long-time behavior.
Here we have assumed that the ultraviolet divergencies
 cannot affect drastically the long-time
behavior.

\section*{Acknowledgments}
We are grateful to V. L'vov and G. Falkovich for valuable remarks.
One of us (I.K.) is grateful to the Foundation of American Physical
Society for partial financial support of this work.

\end{document}